\documentclass[twocolumn,showpacs,aps,prl,superscriptaddress]{revtex4}
\usepackage{graphicx}
\usepackage{dcolumn}
\usepackage{amsmath}
\usepackage{colordvi}
\usepackage{color}
\usepackage{array, longtable}
\usepackage{hhline}
\def\Ecm {\ensuremath{\rm \sqrt{\rm s}}}

\def\epem {\ensuremath{e^+ e^-}}

\def\mevcc {\ensuremath{\rm MeV/c^2}}
\def\piz {\ensuremath{\pi^0}}
\def\ppz {\ensuremath{\pi^0\pi^0}}

\long\def\inst#1{\par\nobreak\kern 4pt\nobreak
    {\it #1}\par\vskip 10pt plus 3pt minus 3pt}

\begin{document}

%



\title{\large \bf
\boldmath
Evidence of the $\epem\to f_1(1285)$ reaction with the CMD-3
detector at VEPP-2000
} 

\author{E.P.\,Solodov}\email[Corresponding author:]{solodov@inp.nsk.su}
\author{R.R.\,Akhmetshin}
\author{A.N.\,Amirkhanov}
\author{A.V.\,Anisenkov}
\author{V.M.\,Aulchenko}
\affiliation{Budker Institute of Nuclear Physics, SB RAS, Novosibirsk, 630090, Russia}
\affiliation{Novosibirsk State University, Novosibirsk, 630090, Russia}
\author{N.S.\,Bashtovoy}
\affiliation{Budker Institute of Nuclear Physics, SB RAS, Novosibirsk, 630090, Russia}
\author{E.V.\,Bedarev}
\affiliation{Budker Institute of Nuclear Physics, SB RAS, Novosibirsk, 630090, Russia}
\affiliation{Novosibirsk State University, Novosibirsk, 630090, Russia}
\author{D.E.\,Berkaev}
\affiliation{Budker Institute of Nuclear Physics, SB RAS, Novosibirsk, 630090, Russia}
\author{A.E.\,Bondar}
\affiliation{Budker Institute of Nuclear Physics, SB RAS, Novosibirsk, 630090, Russia}
\affiliation{Novosibirsk State University, Novosibirsk, 630090, Russia}
\author{A.V.\,Bragin}
\author{D.E.\,Chistyakov}
\author{V.S.\,Denisov}
\author{E.A.\,Eminov}
\affiliation{Budker Institute of Nuclear Physics, SB RAS, Novosibirsk, 630090, Russia}
\author{D.A.\,Epifanov}
\affiliation{Budker Institute of Nuclear Physics, SB RAS, Novosibirsk, 630090, Russia}
\affiliation{Novosibirsk State University, Novosibirsk, 630090, Russia}
\author{L.B.\,Epshteyn}
\affiliation{Budker Institute of Nuclear Physics, SB RAS, Novosibirsk, 630090, Russia}
\affiliation{Novosibirsk State Technical University, Novosibirsk, 630092, Russia}
\author{A.L.\,Erofeev}
\author{G.V.\,Fedotovich}
\affiliation{Budker Institute of Nuclear Physics, SB RAS, Novosibirsk, 630090, Russia}
\affiliation{Novosibirsk State University, Novosibirsk, 630090,Russia}
\author{L.B.\,Fomin}
\affiliation{Budker Institute of Nuclear Physics, SB RAS, Novosibirsk, 630090, Russia}
\author{A.O.\,Gorkovenko}
\affiliation{Budker Institute of Nuclear Physics, SB RAS, Novosibirsk, 630090, Russia}
\affiliation{Novosibirsk State Technical University, Novosibirsk, 630092, Russia}
\author{A.A.\,Grebenuk}
\author{S.S.\,Gribanov}
\affiliation{Budker Institute of Nuclear Physics, SB RAS, Novosibirsk, 630090, Russia}
\affiliation{Novosibirsk State University, Novosibirsk, 630090, Russia}
\author{D.N.\,Grigoriev}
\affiliation{Budker Institute of Nuclear Physics, SB RAS, Novosibirsk, 630090, Russia}
\affiliation{Novosibirsk State Technical University, Novosibirsk, 630092, Russia}
\author{F.V.\,Ignatov}
\author{V.L.\,Ivanov}
\affiliation{Budker Institute of Nuclear Physics, SB RAS, Novosibirsk, 630090, Russia}
\affiliation{Novosibirsk State University, Novosibirsk, 630090, Russia}
\author{A.S.\,Kasaev}
\author{S.V.\,Karpov}
\affiliation{Budker Institute of Nuclear Physics, SB RAS, Novosibirsk, 630090, Russia}
\author{V.F.\,Kazanin}
\author{I.A.\,Koop}
\author{A.A.\,Korobov}
\affiliation{Budker Institute of Nuclear Physics, SB RAS, Novosibirsk, 630090, Russia}
\affiliation{Novosibirsk State University, Novosibirsk, 630090, Russia}
\author{A.N.\,Kozyrev}
\affiliation{Budker Institute of Nuclear Physics, SB RAS, Novosibirsk, 630090, Russia}
\affiliation{Novosibirsk State Technical University, Novosibirsk, 630092, Russia}
\author{P.P.\,Krokovny}
\author{A.S.\,Kuzmin}
\author{T.A.\,Kuznetsov}
\author{I.B.\,Logashenko}
\author{P.A.\,Lukin}
\author{K.Yu.\,Mikhailov}
\author{I.V.\,Obraztsov}
\affiliation{Budker Institute of Nuclear Physics, SB RAS, Novosibirsk, 630090, Russia}
\affiliation{Novosibirsk State University, Novosibirsk, 630090, Russia}
\author{V.G.\,Petrochenko}
\affiliation{Budker Institute of Nuclear Physics, SB RAS, Novosibirsk, 630090, Russia}
\author{N.A.\,Petrov}
\affiliation{Institute for Nuclear Research, RAS, Moscow, 117312, Russia}
\author{A.S.\,Popov}
\affiliation{Budker Institute of Nuclear Physics, SB RAS, Novosibirsk, 630090, Russia}
\affiliation{Novosibirsk State University, Novosibirsk, 630090, Russia}
\author{S.A.\,Rastigeev}
\affiliation{Budker Institute of Nuclear Physics, SB RAS, Novosibirsk, 630090, Russia}
\author{Yu.A.\,Rogovsky}
\affiliation{Budker Institute of Nuclear Physics, SB RAS, Novosibirsk, 630090, Russia}
\affiliation{Novosibirsk State University, Novosibirsk, 630090, Russia}
\author{A.A.\,Ruban}
\affiliation{Budker Institute of Nuclear Physics, SB RAS, Novosibirsk, 630090, Russia}
\author{A.E.\,Ryzhenenkov}
\author{A.V.\,Semenov}
\affiliation{Budker Institute of Nuclear Physics, SB RAS, Novosibirsk, 630090, Russia}
\affiliation{Novosibirsk State University, Novosibirsk, 630090, Russia}
\author{V.E.\,Shebalin}
\affiliation{Budker Institute of Nuclear Physics, SB RAS, Novosibirsk, 630090, Russia}
\affiliation{Novosibirsk State University, Novosibirsk, 630090, Russia}
\author{B.A.\,Shwartz}
\affiliation{Budker Institute of Nuclear Physics, SB RAS, Novosibirsk, 630090, Russia}
\affiliation{Novosibirsk State University, Novosibirsk, 630090, Russia}
\author{D.B.\,Shwartz}
\affiliation{P-cure Ltd, Shilat, 7318800, Israel}
\author{M.V.\,Timoshenko}
\author{V.M.\,Titov}
\affiliation{Budker Institute of Nuclear Physics, SB RAS, Novosibirsk, 630090, Russia}
\author{A.A.\,Talyshev}
\author{S.S.\,Tolmachev}
\affiliation{Budker Institute of Nuclear Physics, SB RAS, Novosibirsk, 630090, Russia}
\affiliation{Novosibirsk State University, Novosibirsk, 630090, Russia}
\author{A.I.\,Vorobiov}
\author{D.S.\,Zhadan}
\affiliation{Budker Institute of Nuclear Physics, SB RAS, Novosibirsk, 630090, Russia}
\author{Yu.V.\,Yudin}
\affiliation{Budker Institute of Nuclear Physics, SB RAS, Novosibirsk, 630090, Russia}
\affiliation{Novosibirsk State University, Novosibirsk, 630090, Russia}
\collaboration{CMD-3 Collaboration}\noaffiliation
%


\begin{abstract}
A search for the $\epem\to f_1(1285)$ reaction  is performed with the
CMD-3 detector at the VEPP-2000  collider.  Data from a scan in the
center-of-mass energy range \Ecm=1.20--1.36 GeV with 109.1\,pb$^{-1}$ of the 
integrated luminosity is used. A dedicated  51.7 pb$^{-1}$ have been collected  at the
$f_1(1285)$ resonance peak energy \Ecm=1282 MeV, where $26.5 \pm 7.4$ events have
been found in the $\epem\to\eta\ppz$ reaction. No indication of the
signal is observed in the non-resonant scan data.
We interpret
the result as a statistical 4.5$\sigma$  evidence of the C-even resonance production in the 
$\epem\to f_1(1285)\to\eta\ppz$
reaction via the two-photon interaction. The  production cross section is found to be
$\sigma(\epem\to f_1) = 0.074 \pm 0.021 \pm 0.010$ nb, and the corresponding 
 branching fraction to the \epem pair is
$B(f_1(1285)\to\epem) = (8.3 \pm 2.3 \pm 1.1)\times 10^{-9}$.
\end{abstract}

\pacs{13.66.Bc, 14.40.Cs, 13.25.Gv, 13.25.Jx, 13.20.Jf}
\maketitle

A single C-even resonance in the \epem~collision can be produced only
via two-photon annihilation, which is suppressed by a factor
$\alpha^2$, fine structure constant squared, comparing with the dominant single photon mechanism. However, 
observation of events in the \epem~collision
for such a resonance provides a significantly better estimate of the
decay rate to \epem~in this "inverse''  reaction in comparison to a
direct decay search.   Many experiments
were performed for the search of  single C-even resonances using
inverse reaction, see for example~\cite{ND,SND1,SND2,SND3,BES1},
but unfortunately only
upper limits were set. For the first time the 
$J^{PC}=1^{++}$ charmonium state $\chi_{c1}(1P)$ has been observed in
the \epem annihilation by the BESIII Collaboration~\cite{BES2} as an
interference pattern in the energy scan, and a branching fraction for
the  $\chi_{c1}(1P)\to\epem$ decay is extracted. 

Axial-vector resonance $f_{1}(1285)$ does not decay into two real photons,
but can be produced via two virtual photons in the \epem~
annihilation. It makes this process very interesting from the
theoretical point of view, in particular for the  understanding of the
light-by-light contribution to the anomalous magnetic moment
of muon $(g-2)$~\cite{dorokhov,kubis}.  

The SND collaboration reported detection of two events, indicated 
the $\epem\to f_1(1285)\to\eta\ppz$ reaction~\cite{sndf1},
corresponding to the branching fraction $B(f_{1}(1285)\to\epem) = (5.1^{+3.7}_{-2.7})\times
10^{-9}$, or to the upper limit  $B(f_{1}(1285)\to\epem) <9.4\times
10^{-9}$ for 90\% CL.
The result is in agreement with theoretical
estimations~\cite{Rud,Milst}, but uncertainties, both in the experiment and
theory, are relatively large.

In this paper we present results of the new search for the
$\epem\to f_1(1285)$ reaction, based on the 109.1\,pb$^{-1}$ integrated luminosity  collected in the
scan of the \Ecm=1200--1360 MeV center-of-mass (c.m.) energy range with 20\,MeV step, and  dedicated 51.7 pb$^{-1}$  at the
c.m.  energy corresponding to the resonance mass $m_{f_{1}}=1281.9\pm0.5~\mevcc$~\cite{pdg}. 
We search  a signal from
the $f_1(1285)$ resonance in the
$f_1\to\eta\ppz$ decay ($B(f_1\to\eta\ppz)= (17.3\pm0.7)\%$)~\cite{pdg}  which is
forbidden for the single photon production. More than 70\% of the
$f_1(1285)\to\eta\ppz$ decays have $a_0(980)\piz$ intermediate state, and
it helps  to identify the reaction.

The general-purpose detector CMD-3 has been described in 
detail elsewhere~\cite{sndcmd3}. The detector tracking system consists of a 
cylindrical drift chamber (DC)~\cite{dc}
placed inside a thin (0.2~X$_0$) superconducting solenoid with a field of 1.3~T.
The barrel liquid-xenon (LXe) calorimeter with a 5.4~X$_0$ thickness
has a fine electrode structure, providing a 1--2 mm spatial resolution 
for photons~\cite{lxe}.
The fine electrode structure allows us to find a conversion point of the 
electromagnetic shower, what significantly improves two photon spatial separation, 
from about 0.4 to 0.15 radians for  the $\gamma\gamma$ open angle, compare to  the
use of the energy deposition location of the clusters.  
The barrel CsI crystal calorimeter is placed outside  the LXe calorimeter 
and  increases the total thickness to   13.5~X$_0$.  The end cap BGO cristal 
calorimeter with a thickness of 13.4~X$_0$ is placed inside the 
solenoid~\cite{cal}.
The luminosity is measured using events of Bhabha scattering 
at large angles with about 1\% accuracy~\cite{lum}. 
The  simulation for the $\epem\to f_1(1285)\to\eta\ppz$ reaction uses a
primary generator with matrix elements, corresponding to the dominant
$f_1 \to a_0(980)\piz$ decay as well as a possible contribution from
the $f_1 \to f_0(500)\eta$ reaction. We also simulate the
$\epem\to\omega\piz$ process for the background study.
The Monte Carlo (MC) simulation of the
detector is based on the GEANT4~\cite{geant4} package, and the
detector conditions are applied to all
generated events which pass
the full reconstruction and selection procedure.
The generators also include a soft photon radiation by the initial 
electron or positron calculated according to Ref.~\cite{kur_fad}. 

For the signal search we  use  the data collected at the expected resonance
maximum \Ecm=1282 MeV with the integrated luminosity of 51.7
pb$^{-1}$. The data from the  energy scan 
in the range \Ecm=1200--1360 MeV with 20 MeV step
with the total  57.4 pb${^{-1}}$  luminosity integral  is used as a
control sample for a background estimate.

\begin{figure}[t]
\begin{center}
\includegraphics[width=0.5\textwidth]{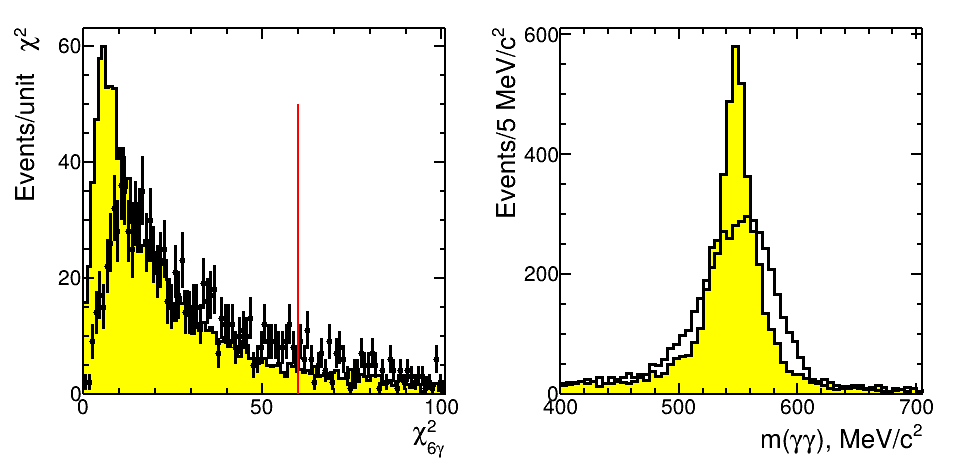}
\put(-180,100){\makebox(0,0)[lb]{\bf(a)}}
\put(-30,100){\makebox(0,0)[lb]{\bf(b)}}
\vspace{-0.5cm}
\caption
{
(a) The 6C-fit $\chi^2_{6\gamma}$  distributions for the simulated
(histogram) and for the experimental (points) events with  two \piz, and two photons
for the $\epem\to\ppz\gamma\gamma$ hypothesis. Line shows
the applied selection.
(b) The simulated two-photon invariant mass distributions
for the $\chi^2_{6\gamma} <60$  selection before (open histogram) and after
(shaded) the kinematic fit.
}
\vspace{-0.7cm}
\label{chi2}
\end{center}
\end{figure}
\begin{figure}[t]
\begin{center}
\includegraphics[width=0.5\textwidth]{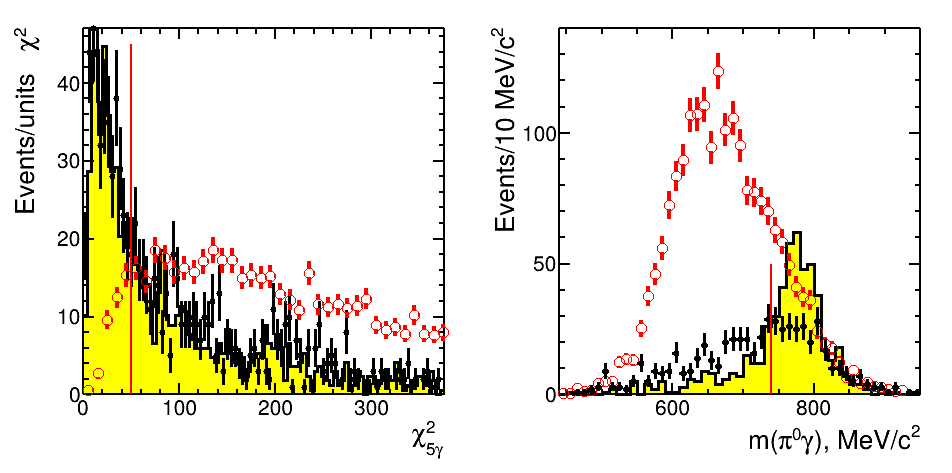}
\put(-180,100){\makebox(0,0)[lb]{\bf(a)}}
\put(-30,100){\makebox(0,0)[lb]{\bf(b)}}
\vspace{-0.5cm}
\caption
{
(a) The  $\chi^2_{5\gamma}$  distributions (the
$\epem\to\ppz\gamma$ hypothesis) for the six-photon events 
with $\chi^{2}_{6\gamma}<60$ for data (points) and
the $\omega\piz$ simulation (shaded histogram, normalized to first 5
bins of data).  The open circles show the
$\chi^{2}_{5\gamma}$ distribution for the $f_1(1285)$ signal simulated
events (not normalized). Line shows the applied selection.
(b) The $\piz\gamma$ invariant mass distributions for events
from (a) after $\chi^2_{5\gamma} >50$   requirement for data (points), simulated $\omega\piz$ events (shaded
histogram), and signal simulated events (open circles, not normalized). Line shows the
applied selection.
}
\vspace{-0.7cm}
\label{chi2_5g}
\end{center}
\end{figure}

We identify the 
$\eta\ppz$ candidate events using the $\eta\to\gamma\gamma$ decay with
six or more (background) photons in the final state. A major background process with a  large
cross section is $\epem\to\omega\piz\to\ppz\gamma$.  This process has five photons
in the final state, but  the  calorimeter detects many soft background
photons with the energy exceeding 25 MeV, therefore it contributes to
the signal sample. We study the $\epem\to\ppz\gamma$
process separately and keep events also with five photons. 
The reconstructed  energy and  angles of each six (five) photon
combination in the event is subject to the 6C kinematic fit~\cite{kinfit} for 
the $\epem\to\ppz\gamma\gamma$ ($\epem\to\ppz\gamma$) hypothesis,
with the total energy and momentum constrained to \Ecm~ and zero,
respectively, and two photon pairs with $| m(\gamma\gamma)- m(\pi^0)|
< 70$~\mevcc~are constrained to the $\pi^0$ mass. All possible photon pair
combinations are tested. As the result of each fit we obtain the improved
photons (photon) and \piz parameters as well as a $\chi^2_{6\gamma}$ ($\chi^2_{5\gamma}$) value.
The parameters of the combination with the smallest $\chi^2_{6\gamma}$ ($\chi^2_{5\gamma}$) value are
retained. Note that the six-photon combination always has the
corresponding five-photon $\chi^2_{5\gamma}$ value.
A shaded histogram in Fig.~\ref{chi2}(a) demonstrates the $\chi^2_{6\gamma} $ distribution
for the $f_1(1285)$ signal simulation, while the points show the
corresponding distribution for data. Figure~\ref{chi2}(b) demonstrates
the improvement for the $\gamma\gamma$ invariant mass
resolution from 25~\mevcc~to about 13~\mevcc.

\begin{figure}[tbh]
  \begin{center}
    \includegraphics[width=0.4\textwidth]{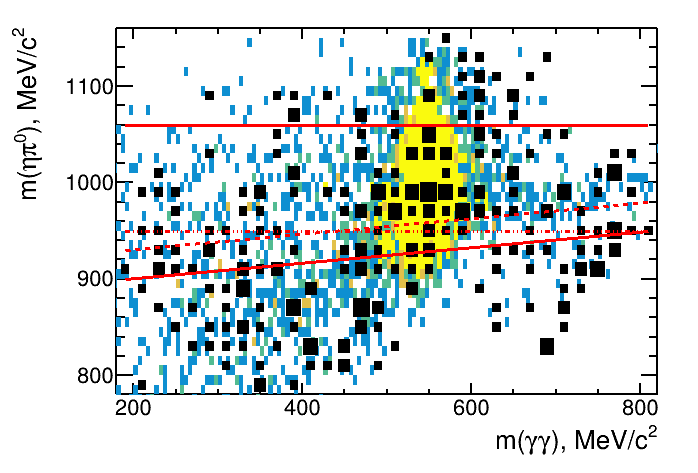}
    \put(-35,105){\makebox(0,0)[lb]{\bf(a)}}\\
    \vspace{-0.3cm}
\includegraphics[width=0.4\textwidth]{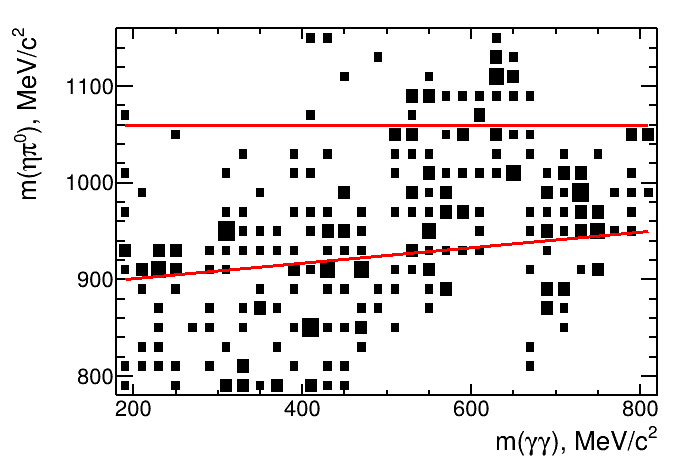}
\put(-35,105){\makebox(0,0)[lb]{\bf(b)}}
\vspace{-0.5cm}
\caption
{(a) The two-dimensional plot of the selected events for the
  $\eta\piz$ invariant mass vs $\gamma\gamma$ invariant mass for 
  \Ecm=1282 MeV for data (box), shown  on top of the simulation. 
  Solid lines show the applied nominal selections. The
   dashed and dash-dotted lines show  the  selections for
the systematic uncertainty estimate.
  (b) The same as (a) for the non-resonant scan data.
  Lines show the applied selections.
 }
\vspace{-0.7cm}
\label{bkgmass}
\end{center}
\end{figure}

For the next analysis step we select events with  $\chi^2_{6\gamma} <
60$ below  the line in Fig.~\ref{chi2}(a).  The
corresponding $\chi^2_{5\gamma}$ distribution for the experimental
events, shown by points  in Fig.~\ref{chi2_5g}(a), demonstrates a
sharp increase to the small values.  It indicates the presence of the $\epem\to\ppz\gamma$
background process when one background photon mimics the signal one.
The  $\chi^2_{5\gamma}$ distribution  for data is 
almost identical to that obtained  
for the simulated $\epem\to\omega\piz$ process, shown by the shaded
histogram in Fig.~\ref{chi2_5g}(a).

The requirement $\chi^2_{5\gamma} > 50$  (line in
Fig.~\ref{chi2_5g}(a)) reduces a large fraction of the  $\omega\piz$ 
background events with the cost of about 10\%  of the signal
events:  the corresponding signal simulated distribution is  shown by
open circles (for demonstration, not in proper normalization). 

For the remaining  events with the two 
$\pi^0$'s and two unconstrained photons  we
calculate a  $\piz\gamma$ invariant mass (take one combination
closest to the $\omega$ mass), and show in
Fig.~\ref{chi2_5g}(b) the distributions for data (points), $\omega\piz$  simulation (shaded
histogram), and signal $f_1(1285)$ simulation (open circles).
A requirement  $m(\piz\gamma) < 740$~\mevcc, shown by line in
Fig.~\ref{chi2_5g}(b),  reduces  the background from
the remaining $\omega(782)\piz$ events with the cost of about 15\% of the simulated signal events.

\begin{figure*}[tbh]
\begin{center}
  \includegraphics[width=0.665\textwidth]{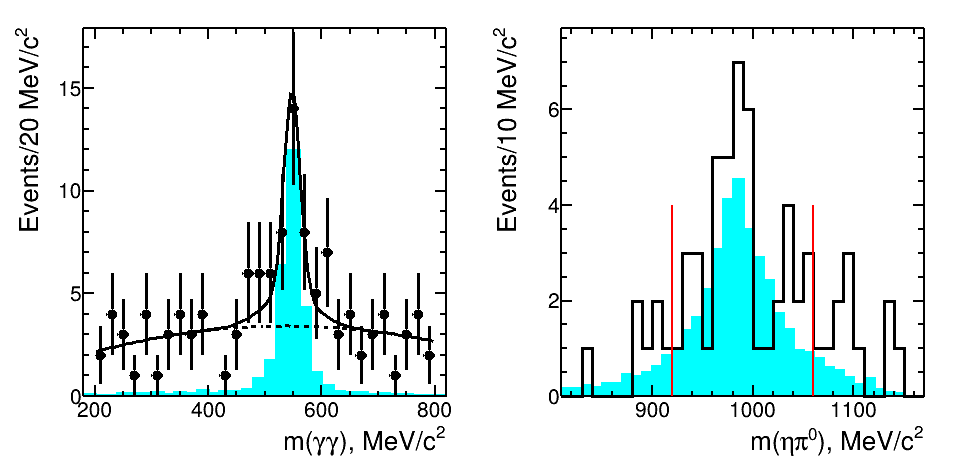}
  \includegraphics[width=0.32\textwidth]{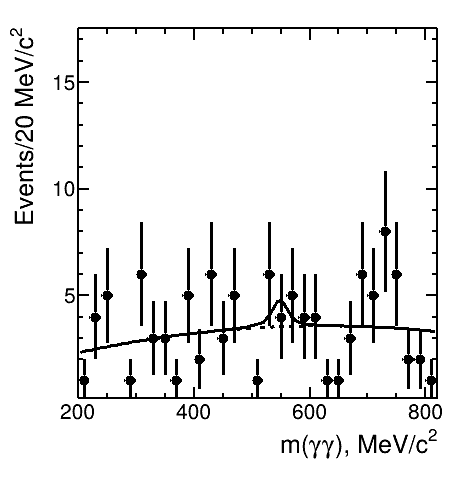}
\put(-370,120){\makebox(0,0)[lb]{\bf(a)}}
\put(-210,120){\makebox(0,0)[lb]{\bf(b)}}
\put(-50,120){\makebox(0,0)[lb]{\bf(c)}}
  \vspace{-0.5cm}
\caption
{(a) The $\gamma\gamma$ invariant mass distribution for events at \Ecm=1282
  MeV selected by solid lines  in Fig.~\ref{bkgmass}(a) with the fit
functions, described in the text. Shaded histogram shows the
simulated distribution, normalized to the observed number of signal events.
(b) The $\eta\piz$ invariant mass selected in the 500--600~\mevcc\,
region of the $\gamma\gamma$ invariant mass in
Fig.~\ref{bkgmass}(a). Shaded histogram shows the simulated
distribution, and lines show the corresponding to lines in Fig.~\ref{bkgmass}(a) selection.
(c) The $\gamma\gamma$ invariant mass distribution for the events from
the non-resonant scan data,
selected and fitted the same way as at \Ecm=1282 MeV. 
}
\vspace{-0.7cm}
\label{Sigf1}
\end{center}
\end{figure*}

Figure~\ref{bkgmass}(a) shows a two-dimensional plot of the $\eta\piz$
invariant mass (one combination closest to the $a_0(980)$ mass is taken)
vs $\gamma\gamma$ invariant mass of the unconstrained photon pair after
applied selections.  A clear
indication of the signal bump at the $a_0(980)$ mass and the
$\eta$ mass is seen. Similar plot for the non-resonant scan data is shown in
Fig.~\ref{bkgmass}(b) with no signal indication.
To maximize the
signal-to-background ratio (FOM = S/sqrt(B))
we select events with m$(\eta\piz) <$ 1060~\mevcc~ and
m$(\eta\piz)>$950+(m$(\gamma\gamma)$ - 800)/12~\mevcc, shown by
solid lines in Fig.~\ref{bkgmass}(a), and make a projection plot of the 
$\gamma\gamma$ invariant mass, shown in
Fig.~\ref{Sigf1}(a). 
This selection makes the background distributions
almost "flat'' and simplifies a signal extraction procedure.
A projection plot of the $\eta\piz$ invariant
mass in the 500--600\,\mevcc~interval of the  $\gamma\gamma$ invariant mass
is shown in Fig.~\ref{Sigf1}(b), demonstrating the $a_0(980)$ resonance signal.

To extract number of signal events we
 fit  the distribution in Fig.~\ref{Sigf1}(a)  with a sum of double
Gaussian function for a signal and 2-nd order polynomial function to
describe a background. All
parameters of the signal function, except 
number of events, are fixed from the fit of the simulated
distribution, shown by a shaded histogram in
Fig.~\ref{Sigf1}(a).
The nominal likelihood fit gives $N_{signal} =  26.5 \pm 7.4$
events with $\chi^2/ndf$ = 21.8/26.  The signal significance is
$4.5\,\sigma$, and is calculated as $\sqrt{2\ln ({\cal L}_{\rm  max}/{\cal
    L}_0)}$, where ${\cal L}_{\rm max}$ and ${\cal L}_0$ are the likelihood values
with and without (shown by a dashed curve in Fig.~\ref{Sigf1}(a)) signal component, respectively.
We use the obtained numbers for the cross section calculation.

A similar fit with the background described by the constant gives
$28.9 \pm 7.3$ events with $\chi^2/ndf$ = 22.7/28 and
$5.4\,\sigma$ significance. We 
cannot prove that the background has uniform distribution, and  the
difference in the number of events with the nominal fit above is used to estimate the
uncertainty from the background subtraction procedure. 

We also use a direct count of the
events in the 500--600\,\mevcc signal region, 41 events,   and use 56
events from the
two intervals, 200--460\,\mevcc~and 640--800\,\mevcc, to estimate
a background level in the  signal mass interval. With the
13.3 events of the estimated background we obtain $27.7 \pm 6.4$ signal events 
associated with the $\epem\to\eta\ppz$ reaction, close to that from
the fits. This approach less depends on the uncertainty in the signal shape but also assumes linear
background shape.

 To be confident that no other processes influence the result, we process
data from the \Ecm = 1200 -- 1360 MeV scan with 20 MeV step 
 with the integrated luminosity of 57.4 pb$^{-1}$ which is comparable with that at
\Ecm=1282 MeV.  All above selections are applied to the scan data, and the
result is shown in Fig.~\ref{bkgmass}(b) and Fig.~\ref{Sigf1}(c).
A background events level for the non-resonant scan data is comparable with that
for the signal region and is almost flat.
No indication of a peaking background is seen.
The fit with the double Gaussian  and 2-nd order polynomial
functions yields $1.8\pm3.5$ signal events, consistent with zero. 

We calculate a  cross section of  the $\epem\to f_1(1285)$
reaction at the resonance maximum using
$$
\sigma_0 = N_{signal}/[L\epsilon(1+\delta)B(f_1 \to\eta\ppz)]
$$
expression, where $L = 51.7$ pb$^{-1}$ is the integrated luminosity,
$\epsilon = 4.95\%$ is the detection efficiency, $(1+\delta) = 0.8$ is
the radiative correction, and $B(f_1 \to\eta\ppz)$ is
the relative decay rate of the $f_1(1285)$ to the detected final state. For the
detection efficiency we use a mixture of the 50000 simulated events in
the $f_1 \to a_0(980)\piz$ decay mode and 20000 events in the $f_1
\to f_0(500)\eta$ mode, approximately corresponding to the
(73$\pm$8)\% dominance of the first one~\cite{pdg}.
Uncertainties in the efficiency and radiative correction are
negligibly small. 
We obtain
 $$
 \sigma_0(\epem\to f_1) = 0.074 \pm 0.021 \pm 0.010 ~{\rm nb}
 $$
 as the peak cross section, and the 
 corresponding branching fraction
 $$
 B(f_1(1285)\to\epem)=\frac{\sigma_0 m^2_{f_{1}}}{12\pi C}=
(8.3 \pm 2.3 \pm 1.1)\times 10^{-9},
 $$
where $C = 3.8938\times 10^{11} $ MeV$^2$nb  is the conversion
constant.

 Systematic uncertainties, shown as the second errors above,  are
 dominated by the uncertainty in the background 
 subtraction procedure, 1-2 events, and the uncertainties from variations
 of the selection criteria  in the 2D plot,  shown by the
 dashed and dash-dotted lines in Fig.~\ref{bkgmass}(a), 10\% in total.
 The uncertainty in the photon detection efficiency and from a possible wrong \piz combination
 is estimated by the comparison of the 
 $\epem\to\omega\piz\to\ppz\gamma$  cross section, events are extracted and
 normalized in parallel to the six-photon study, 
 with the other measurement from Ref.~\cite{snd5g}. This uncertainty
 does not exceed 5\%.
 The uncertainty
 in the $f_1(1285)$ decay modes~\cite{pdg} contributes about 5\%.

 In summary: we perform a search for the $\epem\to\eta\ppz$ reaction with
the CMD-3 detector at the VEPP-2000 collider. At the expected
$f_1(1285)$ resonance maximum with the 51.7 pb$^{-1}$ of the integrated
luminosity $26.5 \pm 7.4$ events have been found
corresponding  $4.5~\sigma$ significance of the signal.   In the
non-resonant energy scan  with the 57.4
pb$^{-1}$ of the integrated luminosity, no significant signal  is 
observed. The calculated cross section $\sigma_0(\epem\to f_1) =
0.074 \pm 0.021 \pm 0.010$ nb 
 is in agreement with that presented by the SND experiment~\cite{sndf1}.
 The branching fraction of the $f_1(1285)$ to the \epem
 pair is found to be
 $B(f_1(1285)\to\epem)=(8.3 \pm 2.3 \pm 1.1)\times10^{-9}$.
 The result is not in contradiction with the theoretical estimate in
 Ref.~\cite{Milst}, but seems to be in tension with the prediction in Ref.~\cite{kubis}. 
 The systematic uncertainties allow us conservatively to interpret our 
 result as the first evidence of the direct C-even resonance production
 in the \epem collision.

\end{document}